\documentclass[conference]{IEEEtran}
\IEEEoverridecommandlockouts

\usepackage{booktabs}
\usepackage{tabularx}
\usepackage[sorting=none]{biblatex}
\usepackage{amsmath,amssymb,amsfonts}
\usepackage{algorithmic}
\usepackage{graphicx}
\usepackage{textcomp}
\usepackage{xcolor}

\def\BibTeX{{\rm B\kern-.05em{\sc i\kern-.025em b}\kern-.08em
    T\kern-.1667em\lower.7ex\hbox{E}\kern-.125emX}}

\begin{document}

\title{Evaluating Multimodal Steganalysis for Split-Payload Audiovisual Steganography}

\author{%
  Prateek~Paudel\textsuperscript{\dag *}, \quad 
    Nitin~Jha\textsuperscript{\dag}, \quad
  Abhishek~Parakh\textsuperscript{\dag}, \quad \\
  \textsuperscript{\dag}Kennesaw State University, Marietta, GA, USA\\[1ex]

  \textsuperscript{*}\texttt{Corresponding Author: ppaudel@students.kennesaw.edu}\\

}

\maketitle

\begin{abstract} 
The aim of steganography is to hide secret information inside ordinary media so that the existence of communication is hidden rather than encrypted. In audiovisual context, the availability of audio and video streams creates an opportunity to split a payload across these two modes thus, reducing the embedding burden on any single carrier. This paper evaluates whether such split-payload audiovisual steganography can help evade unimodal and multimodal steganalysis under synchronized and asynchronous embedding settings. We create audiovisual samples where the hidden message is divided between the audio and video tracks, and then test how well different detectors can identify them. The single mode detectors performs close to random guessing, thus showing the benefit of this hiding mechanism, while the multimodal model initially appears more effective. However, further checks show that this improvement mostly comes from the video stream, not from a true combined audio-video signal. Overall, the results suggest that splitting the payload across modalities can make detection harder, but multimodal detectors must be evaluated carefully to ensure they are learning the intended signal.

\end{abstract}


\section{Introduction}

Steganography is one of the oldest forms of covert communication, with roots dating back to the fifteenth century. Its goal is to conceal a secret message within another seemingly ordinary medium, commonly called the cover, so that the very existence of the hidden communication is obscured. This distinguishes steganography from cryptography: whereas cryptography protects the content of a message by encrypting it, steganography seeks to hide the presence of the message itself. As a result, the cover medium may remain publicly visible, while only the intended recipient recognizes that secret information is embedded and can recover it.

\begin{figure*}[!htpb]
    \centering
    \includegraphics[width=\textwidth]{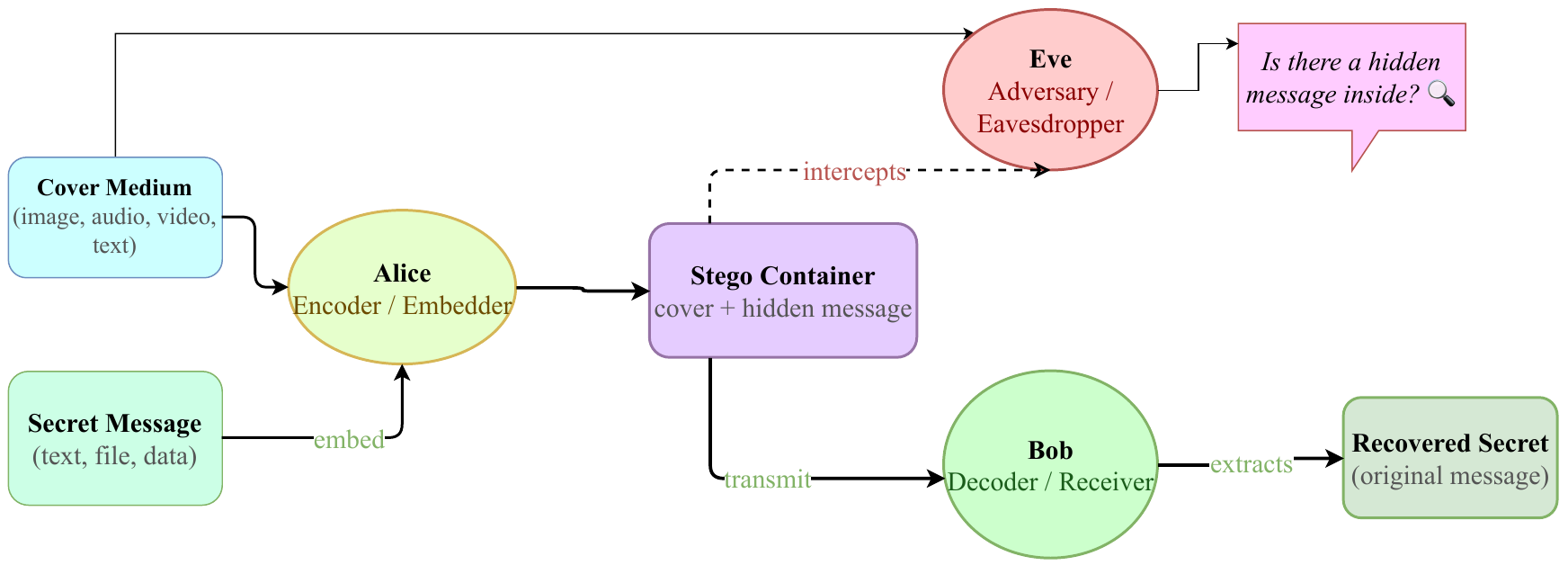}
    \caption{Schematic representation of the general working of a steganography algorithm for hiding secret messages within a cover medium.}
    \label{fig:stego}
\end{figure*}

In the classic setup, there are three parties: Alice, Bob, and Eve. Alice embeds a secret message into a cover image to produce a \textit{container}, which is then sent to Bob. Eve plays the role of the adversarial steganalyzer and attempts to determine whether a given clip is an ordinary cover or a container with a hidden payload. Alice succeeds ideally when Bob can decode the embedded secret with high reliability, while Eve's detection accuracy is no better than chance, that is, around 50\%. This mirrors the adversarial objective of making container objects statistically indistinguishable from genuine covers. Figure~\ref{fig:stego} illustrates the general structure of a steganography algorithm.

Image steganography has attracted significant research attention for many years~\cite{karim2011new, islam2014edge, kombrink2024image}. A common measure of embedding load is bits per pixel (bpp), representing the average number of secret bits hidden in each pixel of the cover image. Traditional methods typically operate below 1~bpp to preserve visual quality and avoid detection. More recent work has pushed toward high-capacity steganography, with some methods hiding an entire full-resolution color image inside another of the same spatial dimensions, corresponding to a payload of up to 24~bpp. The challenge in this regime is balancing large embedding volume against imperceptibility and resistance to detection. Video steganography, which involves embedding secret information into a cover video, has received growing attention because videos offer both large carrier capacity and rich spatial-temporal redundancy~\cite{weng2019high}. Compared to image steganography, the video setting is more complex since it requires preserving not only the visual quality of individual frames but also temporal consistency across them. More recent methods do not treat a video as a simple collection of independent frames. Instead, they take advantage of motion, temporal dependencies, edge information, and learned high-level features to improve both imperceptibility and robustness. Some deep learning approaches use selective frame embedding and sparse modifications to harden detection, while others focus on preserving high-resolution quality through temporal guidance, edge-aware strategies, or semantic-level hiding within generative video frameworks~\cite{chen2024deep}.

In this work, we study a steganographic scheme in which a secret message is split between the audio and video tracks of a single clip, with each track carrying only a fraction of the total payload. The central question is whether this split-payload design is harder to detect than conventional single-medium embedding, and whether a multimodal detector can recover a reliable signal when neither track alone provides one. We construct samples under several related embedding modes and evaluate all of them against a set of unimodal and multimodal detectors to map out the conditions under which the approach succeeds or fails. The results support a multimodal detection advantage under synchronized split-payload conditions, while also revealing open questions that motivate future ablation work.

This paper is organized as follows. Section~\ref{Sec:related} reviews prior work on adaptive embedding, deep steganalysis, audio steganalysis, video steganography, and multimodal representation learning. Section~\ref{Sec:methodology} describes the experimental setup. Section~\ref{Sec:Results} presents and interprets the results. Section~\ref{Sec:conclusions} summarizes the findings and outlines directions for future work.

\section{Related Work}
\label{Sec:related}

This work touches on five areas of prior research: adaptive embedding, deep image steganalysis, audio steganalysis, video steganography, and multimodal representation learning.

Work on the embedding side of steganography has gradually shifted away from simple substitution-based methods toward adaptive schemes that account for perceptual distortion. A key example is the UNIWARD framework, which showed that modifications should be placed in textured or complex regions where they are least noticeable and hardest to detect~\cite{holub2014universal}. This idea directly motivates the content-adaptive cost functions used in our video and audio embedding pipelines.

On the detection side, deep learning has largely replaced feature-engineering approaches in image steganalysis. Xu et al.\ showed that CNNs designed to capture weak residual artifacts, rather than semantic image content, can outperform traditional pipelines~\cite{xu2016structural}. Ye et al.\ extended this by showing that emphasizing high-pass residuals and using task-specific nonlinearities can further improve detection accuracy. Boroumand et al.\ consolidated these ideas in SRNet, a residual architecture that remains one of the most widely used baselines for spatial steganalysis~\cite{Boroumand2019SRNet}. Because our video detector uses an SRNet-style encoder to analyze frame-level embedding artifacts, these works provide the closest foundation for the visual component of our design.

Audio steganalysis has received considerably less attention. One relevant contribution is the CNN-LSTM framework proposed by Yang et al.\ for VoIP steganalysis, which pairs high-pass residual extraction with temporal sequence modeling~\cite{yang2019steganalysis}. Their results suggest that weak audio embedding artifacts are better captured when local convolutional filtering is combined with recurrent modeling over time, an observation that directly informs our audio encoder design.

In video steganography, Weng et al.\ showed that temporal residual modeling can support high-capacity embedding while preserving inter-frame consistency, and argued that treating a video as an independent collection of images ignores important temporal structure~\cite{weng2019high}. Hu et al.\ introduced StegaVideo, which uses temporal and edge guidance to improve robustness and visual quality in high-resolution video~\cite{hu2024stegavideo}. Mao et al.\ proposed hiding information within the semantic feature space during video editing, using a generative framework to improve imperceptibility~\cite{mao2024covert}. Together, these works highlight the importance of temporal consistency and motion-aware design, both of which are central to our synchronized slice-based framework.

Our multimodal detector design is also informed by progress in cross-modal representation learning. AV-HuBERT demonstrated that jointly modeling temporally aligned audio and video streams produces strong multimodal representations~\cite{shi2022learning}, while SimCLR showed that contrastive learning can effectively separate aligned and misaligned samples in a shared latent space~\cite{chen2020simple}. These ideas motivate the design philosophy of the cross-attention detector used in this study.

\section{Methodology}
\label{Sec:methodology}

This section describes the experimental framework used to evaluate Split-payload audiovisual steganography and its detectability. The central hypothesis is that distributing a secret message across two modalities at individually sub-threshold rates suppresses unimodal detection. We further investigate whether multimodal architectures can recover a joint signal under these conditions, or whether the per-modality rates are low enough to defeat cross-modal analysis as well.

\subsection{Problem Setup}

Let $(\mathbf{V}, \mathbf{A})$ denote an audiovisual clip, where
$\mathbf{V} \in \{0,\ldots,255\}^{T \times H \times W}$ is a grayscale video sequence and $\mathbf{A} \in \mathbb{Z}^{N_s}$ is a monaural PCM waveform. Detection is framed as a binary hypothesis test:
\begin{equation}
    H_0 : (\mathbf{V}, \mathbf{A}) \sim P_{\text{cover}},
    \quad
    H_1 : (\mathbf{V}, \mathbf{A}) \sim P_{\text{stego}}.
\end{equation}

In the split-payload setting, neither modality carries the full message. Each track is embedded at a rate calibrated to fall below the detection threshold of its corresponding unimodal classifier. In the
\emph{synchronized} variant, both embedders share a common per-slice seed, introducing a statistical dependency between the modification patterns of the two tracks:
\begin{equation}
    I(\phi^v_k;\, \phi^a_k \mid s_k) > 0 \quad \text{(sync)}.
\end{equation} This dependency is absent from the cover material. Whether a multimodal detector can exploit it when neither marginal signal is individually reliable is the question the experiments address.

\begin{figure*}[!htpb]
    \centering
    \includegraphics[width=\linewidth]{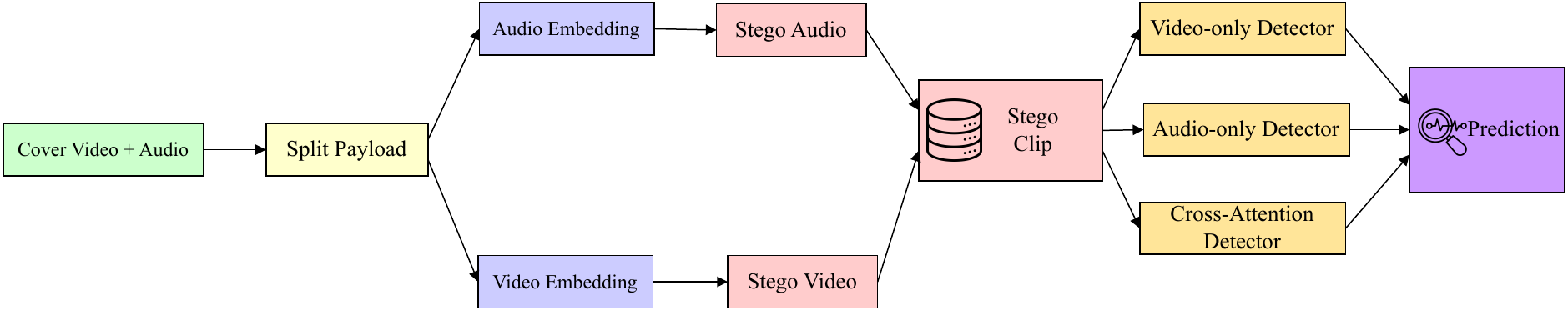}
    \caption{Schematic diagram of the experimental pipeline.}
    \label{fig:stegoflow}
\end{figure*}
\subsection{Temporal Slicing and Synchronization}

Both modalities are divided into $K$ contiguous temporal slices before embedding. A clip-level master seed produces a per-slice seed vector
$\mathbf{s} = (s_0, \ldots, s_{K-1})$ and a payload-rate modulation vector $\boldsymbol{\lambda}$, drawn uniformly and clipped to a stable range. In sync mode, both the video and audio embedders at slice $k$
are initialized with $s_k$, making the modification locations correlated. In \texttt{async} mode, the audio embedder receives independently drawn seeds while the video embedder retains $s_k$:
\begin{equation}
    \phi^v_k \perp \phi^a_k \quad \text{(\texttt{async})}.
\end{equation} Because $\boldsymbol{\lambda}$ is computed before the seed streams diverge, per-slice payload rates are identical in both modes. Any difference in detector performance between sync and \texttt{\texttt{async}}, therefore, reflects the presence or absence of cross-modal seed correlation, not a difference in embedding volume.

\subsection{Embedding Pipeline}

\paragraph{Video.} Video frames are embedded using the WOW algorithm~\cite{holub2014universal}, a content-adaptive scheme that concentrates modifications in textured regions. The per-pixel cost is
\begin{equation}
    \rho_{ij} = \left(\sum_{d \in \mathcal{D}}
    |\mathbf{h}_d \ast \mathbf{F}|_{ij}\right)^{-1} + \epsilon,
\end{equation} where $\mathcal{D}$ denotes a set of directional high-pass filters. WOW is chosen because its content-adaptive placement raises the bar for unimodal frame-level detectors, reducing the risk that any observed multimodal advantage stems from trivially detectable visual artifacts.

\paragraph{Audio.} Audio samples are embedded using a syndrome-trellis code (STC) with energy-based per-sample costs:
\begin{equation}
    \rho^a_n = \frac{1}{\mathrm{RMS}(n) + \epsilon},
    \quad
    \mathrm{RMS}(n) = \sqrt{\frac{1}{W}\sum_{l=0}^{W-1} a_{n-l}^2}.
\end{equation} This assigns high cost to silent regions and low cost to energetic speech, exploiting psychoacoustic masking analogously to WOW's spatial cost assignment. The resulting PESQ MOS score exceeds 4.49 at 0.25 bps, confirming perceptual transparency.

\paragraph{Distortion equalization.} To prevent the detection problem from reducing to unimodal analysis of a single heavily loaded track, the audio payload rate $\beta_0$ is calibrated via binary search so that perceptual distortion in each modality is comparable:
\begin{equation}
    |\Delta_v(\alpha_0) - \Delta_a(\beta_0)| < \tau,
\end{equation} where $\Delta_v = 1 - \mathrm{SSIM}$ is video distortion and
$\Delta_a$ is PESQ-normalized audio distortion.

\subsection{Dataset Construction}

All experiments use clips from the RAVDESS corpus~\cite{livingstone2018ryerson}, resampled to a standardized frame count, spatial resolution, and audio sampling rate before embedding. Dataset splits are actor-disjoint: test actors (22--24) are never seen during training or validation. This design guards against identity leakage, though the small number of test actors remains a limitation that we address in Section~\ref{sec:interpretation}.

Four embedding modes implement a partial factorial design over seed assignment and payload distribution. Table~\ref{tab:modes} summarizes their roles.

\begin{table}[!htpb]
\centering
\caption{Embedding modes and their experimental roles.}
\label{tab:modes}
\begin{tabularx}{\columnwidth}{|l|l|l|X|}
\toprule
\textbf{Mode} & \textbf{Payload} & \textbf{Seed} & \textbf{Role} \\
\midrule 
sync         & Balanced       & Shared/slice  & Primary cross-modal test   \\
\texttt{async}        & Balanced       & Independent   & Synchronization ablation   \\
audio\_heavy & Audio-focused  & Shared/slice  & Audio unimodal baseline    \\
video\_heavy & Video-focused  & Shared/slice  & Video unimodal baseline    \\
\bottomrule
\end{tabularx}
\end{table}

\subsection{Detector Architectures}

All detectors receive $K$ video slices and $K$ audio slices and return a binary cover/stego prediction. The differences concern only how each detector encodes and combines the two streams.

\paragraph{Video encoder (SRNet-style).} A convolutional stem with absolute-value activation extracts residuals that are symmetric with respect to $\pm 1$ LSB modifications. Subsequent residual blocks operate at full spatial resolution before a pooling stage reduces each frame to a fixed-length vector. Spatial down-sampling uses nearest-neighbor interpolation rather than bilinear, since bilinear averaging attenuates $\pm 1$ residuals before they reach the nonlinearity~\cite{Boroumand2019SRNet}.

\paragraph{Audio encoder (CNN-LSTM).} Normalized PCM audio is processed with a dual-polarity high-pass filter bank followed by absolute-value activation. Multiscale convolutional branches extract residuals at different temporal resolutions, and a bidirectional LSTM captures the sequential structure within each slice~\cite{yang2019steganalysis}. 
\paragraph{Shared temporal module.} Both unimodal and multimodal detectors pass their slice-level embeddings through a bidirectional LSTM over the $K$-slice sequence, followed by multi-head attention pooling and a classification MLP. Attention pooling is preferred over mean pooling because silent segments contain no audio modifications, and some temporal positions may carry more information than others. Using an identical temporal module across all detectors ensure that performance differences reflect fusion strategy, not temporal modeling capacity.

\paragraph{Cross-attention fusion.} The primary multimodal detector consists of three bidirectional cross-attention transformer layers. Each layer applies self-attention within each modality first, then cross-attention between video and audio sequences. A shared positional embedding encourages corresponding slice positions to attend to one another. The final classification head receives the concatenation of mean-pooled video and audio representations.

\subsection{Post-Hoc Ablation Protocol}

All ablation studies operate on frozen checkpoints with no retraining.

\paragraph{Modality masking.} The audio-masking ablation replaces all audio inp t with zeros while preserving the tensor shape, so the forward pass proceeds normally but receives no audio content. The video-masking ablation applies the same treatment to the video stream. If the model has learned a genuine multimodal signal, removing either input should degrade performance.

\paragraph{Same-label shuffled pairing.} To test whether the detector relies on within-clip temporal correspondence between audio and video, we replace each sample's audio with audio from a \emph{different} clip carrying the \emph{same} cover/stego label. This preserves label consistency while destroying any cross-modal temporal alignment. Four stride offsets are evaluated and averaged to guard against accidentally correlated pairings.

\subsection{Training Details}

All models are trained with AdamW, cosine learning rate decay, and a linear warmup period. Cross-entropy loss with label smoothing ($\epsilon = 0.1$) is used throughout. Warmup is particularly important in the multimodal setting, where random initialization produces uninformative features that can interfere destructively across branches if the full learning rate is applied before either stream has stabilized. Early stopping is applied on validation $P_E$, and the best-performing checkpoint is retained for test evaluation.

Two label-preserving augmentations are applied during training. Brightness jitter scales the pixel array by a uniform factor ($\pm 5\%$), which does not alter WOW's high-pass filter responses and therefore preserves the embedding selection criteria. Low-magnitude Gaussian noise is added to the audio signal to reduce overfitting to precise waveform values in the limited training corpus. Horizontal flipping is excluded because WOW's diagonal filter kernel is not symmetric under reflection. Temporal slice reordering is also excluded, as it would destroy the synchronization structure the detectors are designed to exploit.

\subsection{Evaluation Metrics}

The primary metric is the probability of error:

\begin{equation}
    P_E = \tfrac{1}{2}(P_{FA} + P_{MD}),
\end{equation}

where $P_{FA}$ is the false alarm rate and $P_{MD}$ is the miss-detection rate. $P_E = 0.5$ corresponds to random guessing. Accuracy and AUC-ROC are reported alongside $P_E$.

\section{Results}
\label{Sec:Results}

We consider two major questions here. First, is the proposed cross-attention detector capable of detecting steganographic artifacts in the split-payload setting? Second, even if aggregate accuracy is misleading, what is the actual detector's behavior?
Both these issues can be addressed with the help of the ablation protocol discussed in Sec.\ref{Sec:methodology}.

\subsection{Multimodal Detection Performance}

Table~\ref{tab:main_results} shows the test set performance of detector-embedding combinations used in the experiment. Cross-attention detector demonstrates 93.62\% of test accuracy ($P_E = 0.064$) on synchronized split-payload embedding, while both unimodal baselines have chance-level performance ($P_E = 0.50$). Unimodal baseline detectors are still at chance level on the
\texttt{async} condition. This result confirms the assumption that split-payload embedding prevents a single-modality detector from learning the embedding, regardless of seed selection. On \texttt{async} embedding, cross-attention detector accuracy drops to 82.75\% ($P_E = 0.173$).

\begin{table}[t]
\centering
\caption{Test set performance of the detectors with the split-payload embedding, in synchronized and \texttt{async} modes. Note that only payload rate differs between modes, the rest of the setup is unchanged. All four unimodal baseline setups are at chance level.}
\label{tab:main_results}
\begin{tabular}{|llccc|}
\hline Detector        & Mode  & Acc.    & $P_E$   & AUC    \\
\hline Cross-attention & Sync  & 93.62\% &  6.37\% & 0.906 \\
Cross-attention & \texttt{async} & 82.75\% & 17.25\% & 0.995 \\
Video-only      & Sync  & 50.00\% & 50.00\% & 0.499 \\
Video-only      & \texttt{async} & 50.00\% & 50.00\% & 0.500 \\
Audio-only      & Sync  & 50.00\% & 50.00\% & 0.500 \\
Audio-only      & \texttt{async} & 50.00\% & 50.00\% & 0.500 \\
\hline
\end{tabular}
\end{table}

These results may imply that cross-attention detector outperforms the other two setups in terms of test accuracy; there is a synchronization gap and hence some advantage to working in the multimodal environment. However, further analysis reveals what signals are actually learned by the model.


\subsection{Modality Contribution Ablation}

We consider results of masking audio and video inputs and shuffling audio-video pairing.

\begin{table}[t]
\centering
\caption{Test set performance with various input modification.
$\Delta P_E$ is the difference from the baseline setup. Shuffled audio-video pairs were produced using four strides, each having $\sigma < 0.001$.}
\label{tab:modality_ablation}
\begin{tabular}{|lcccc|}
\hline
\textbf{Condition} & \textbf{Acc.} & $\mathbf{P_E}$ &
$\mathbf{\Delta P_E}$ & \textbf{AUC} \\
\hline
\multicolumn{5}{l}{\textit{Synchronized embedding}} \\
Baseline            & 93.62\% & 0.064 & ---      & 0.906 \\
Audio zeroed        & 93.50\% & 0.065 & +0.001   & 0.904 \\
Video zeroed        & 50.00\% & 0.500 & +0.436   & 0.500 \\
Audio shuffled      & 93.60\% & 0.064 & $<$0.001 & 0.905 \\
\hline
\multicolumn{5}{l}{\textit{\texttt{async}hronous embedding}} \\
Baseline            & 82.75\% & 0.173 & ---      & 0.995 \\
Audio zeroed        & 82.75\% & 0.173 & +0.000   & 0.995 \\
Video zeroed        & 50.00\% & 0.500 & +0.328   & 0.500 \\
Audio shuffled      & 82.75\% & 0.173 & +0.000   & 0.995 \\
\hline
\end{tabular}
\end{table}

We can draw several conclusions from this table. First, removing the audio stream changes test performance by less than 0.2 percentage points. For \texttt{async} mode, the effect is exactly zero. Second, removal of the video stream drops test accuracy to chance level $P_E = 0.50$. Third, shuffling audio-video pair does not affect classification performance in either mode.In this way, the role of each modality becomes quite obvious. Cross-attention detector essentially acts as the video-only one; it neither uses audio data nor exploits the cross-modal temporal correspondence. Three transformer layers which connect two branches play merely structural role.

\subsection{Interpretation}
\label{sec:interpretation}

A combination of masking of each modality and shuffling audio-video pairs leads to the conclusion that cross-attention detector does not detect the hidden message in the split-payload condition. Its high test set accuracy of 93.62\% is explained solely by video modality usage. Disaggregation of test performance by speaker identity reveals that all 153 false negatives in the synchronized embedding belong to a single speaker, while the remaining two speakers are classified without error. This clustering of test mistakes may indicate speaker identity-
dependent shortcut learning rather than actual signal detection. Moreover, such a concentration of false negatives also makes us doubt about the appropriateness of performing evaluation on such small test partitions with only three speakers.

There are two implications of this result for the split-payload evasion approach. First, the approach itself appears to be correct: embedding at low rates in multiple modalities ($0.25\,bpp$, $0.08\,bps$) prevents multimodal detector from learning the combined embedding artifact, even if it uses explicit cross-modality attention. Test accuracy of both audio and video unimodal baselines is chance level in all four detector configurations (Table~\ref{tab:main_results}). At the same time, STC embedding produces perceptually transparent result (PESQ MOS greater than 4.49), without creating additional cross-modal signal that could be detected by cross-attention model.

Second, this research suggests an important aspect of methodological issues in multimodal steganalysis. Small-speaker corpus like RAVDESS contains only a limited number of different speakers. Features of their identity become correlated with the actor-disjoint split. Any detector that is able to capture identity-specific visual features would exhibit good performance on overall metrics, but would not actually learn anything about the embedding artifact. The suggested approach, combining modality masking with disrupting temporal alignment, offers a way to detect this problem. We encourage researchers to adopt this practice in their works.

\section{Conclusions}
\label{Sec:conclusions}

This paper investigates a steganographic approach where secret message is distributed across the audio and video tracks within a single clip at sub-threshold payload rates ($0.25\,bpp$ video, $0.08\,bps$
audio). Neither the video nor audio unimodal detector was able to detect embedded payloads in both synchronized and \texttt{async} conditions (test accuracy $P_E = 0.50$). At the same time, cross-attention multimodal detector demonstrated 93.62\% test accuracy. However, further analysis showed that this result cannot be explained by detection of steganographic artifacts.

We saw zero impact on test accuracy when zeroing audio stream, complete drop of accuracy when removing video modality and Lack of impact caused by shuffling audio and video pairings. We also saw disaggregation of predictions according to speaker identity shows that all 153 false negatives of the synchronous embedding belong to a single test speaker.

These findings confirm the correctness of the split-payload evasion strategy, while emphasizing the importance of ablation-based analysis in multimodal steganalysis experiments. Lack of audio modality impact, lack of sensitivity to shuffling of pairs across samples, and reliance solely on video branch, serve as strong evidences for difficulty of cross-modal embedding detection. Further research should extend the experiment to multi-speaker datasets and explore the possibility of detecting cross-modal signal at higher payload rates.

\printbibliography

\end{document}